\documentclass{moriond}

\usepackage{lipsum}  
\usepackage{xspace}

\def\be{\begin{equation}}
\def\ee{\end{equation}}
\def\bea{\begin{eqnarray}}
\def\eea{\end{eqnarray}}

\newcommand{\Euclid}{\textit{Euclid}\xspace}
\newcommand{\sfont}[1]{{\scriptscriptstyle\rm #1}}
\newcommand{\IE}{\ensuremath{I_\sfont{E}}}
\newcommand{\YE}{\ensuremath{Y_\sfont{E}}}
\newcommand{\JE}{\ensuremath{J_\sfont{E}}}
\newcommand{\HE}{\ensuremath{H_\sfont{E}}}

\newcommand*{\arcsecf}{\hbox{$.\!\!^{\prime\prime}$}}

\newcommand*{\AckInstitutions}{a number of agencies and
  institutes that have supported the development of \Euclid, in
  particular the Agenzia Spaziale
  Italiana, the Belgian Science Policy, the Canadian Euclid
  Consortium, the French Centre National d'Etudes Spatiales, the
  Deutsches Zentrum f\"ur Luft- und Raumfahrt, the Danish Space
  Research Institute, the Funda\c{c}\~{a}o para a Ci\^{e}ncia e a
  Tecnologia, the Hungarian Academy of Sciences, the Ministerio de Ciencia,
  Innovaci\'{o}n y Universidades, the National Aeronautics and Space
  Administration, the National Astronomical Observatory of Japan,
  the Netherlandse Onderzoekschool Voor Astronomie, the Norwegian
  Space Agency, the Research Council of Finland,
  the Romanian Space Agency, the State Secretariat
  for Education, Research and Innovation (SERI) at the Swiss
  Space Office (SSO), and the United Kingdom Space Agency. A
  complete and detailed list is available on the \Euclid\ web site
  (\texttt{http://www.euclid-ec.org}).}
\newcommand{\AckEC}{The Euclid Consortium acknowledges the European
  Space Agency and \AckInstitutions}

\newcommand{\Photo}{\includegraphics[height=35mm]{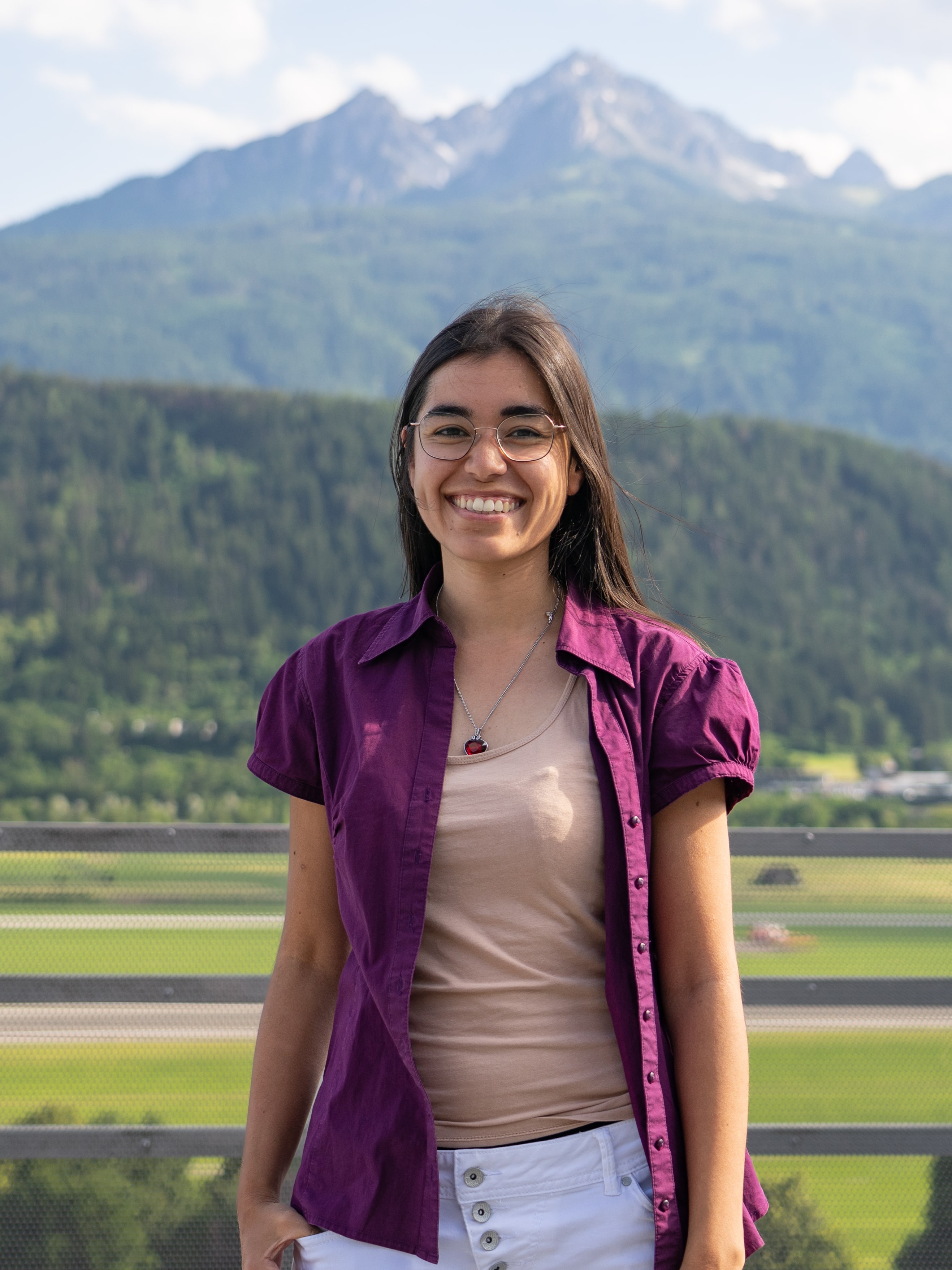}}

\begin{document}
\vspace*{4cm}
\title{\textit{Euclid} -- the Dark Universe detective}

\author{ L.~Linke on behalf of the Euclid Consortium }

\address{Universit\"at Innsbruck, Institut f\"ur Astro- und Teilchenphysik, Technikerstr. 25/8, 6020 Innsbruck, Austria}

\maketitle\abstracts{\Euclid is a recently launched medium-class mission by the European Space Agency (ESA) designed to measure cosmological parameters, test the cosmological standard model, and explore the nature of dark matter and dark energy. To this end, \Euclid conducts a survey of up to $14\,000\,\mathrm{deg}^2$ of the extra-galactic sky and obtains optical and near-infrared photometric measurements for more than a billion galaxies as well as near-infrared slitless spectroscopy for more than 35 million galaxies. These observations will be used to estimate galaxy clustering and cosmic shear. It is expected that \Euclid will achieve percent-level constraints on the Dark Energy equation of state parameter. The survey will also be exploited with a range of other cosmological probes and prove revolutionary for non-cosmological science.}

\section{Introduction}

The cosmological standard model, $\Lambda$CDM, is a remarkable success story, able to describe a wide range of observations from the early Universe and the cosmic microwave background (CMB) to today's large-scale structure (LSS) with just six parameters\,\cite{Planck2020_VI}\;\cite{DES2023}\;\cite{Troester2021}. However, a range of issues have emerged with this standard model. The first and most pressing is the nature of the dark universe. As demonstrated by many independent observations, e.g., the CMB, supernovae, or the LSS, the energy-matter budget of the Universe is dominated by dark matter and dark energy, both of which cannot be described by the standard particle physics and are fundamental puzzles to modern physics.

Another emerging issue is the difference in inferred parameter values from probes of the early and late Universe. For example, measurements of the CMB\,\cite{Planck2020_VI}  predict, under the assumption of $\Lambda$CDM, a Hubble constant of $H_0 = 67.5 \pm 0.5 \, \mathrm{km}\,\mathrm{s}^{-1}\,\mathrm{Mpc}$, while direct measurements via the distances and redshifts of cepheids and supernovae Type Ia\,\cite{Riess2022} suggest $H_0 = 73.04 \pm 1.04 \,\mathrm{km}\,\mathrm{s}^{-1}\,\mathrm{Mpc}$ -- a $5\sigma$ tension. A similar, though less significant, tension is seen for the clustering parameter $S_8$, for which measurements from the LSS at late times predict lower values than inferred from the CMB\,\cite{Abdalla2022}.

To confirm or reject these tensions and to study the nature of dark matter and dark energy, new observations are needed. Observations of the CMB are not optimal for this quest, as the CMB is mainly sensitive to the properties of the Universe at recombination when dark energy was negligible. Instead, the cosmic expansion history and the growth of the LSS provide more information on the influence of the dark universe. Accessing this information is the principal goal of the recently started \Euclid mission\,\cite{Laureijs2011}, which conducts space-based surveys of the extragalactic sky and infers cosmological information from two probes of the LSS.

The first probe is the three-dimensional clustering of galaxy positions. These galaxies trace the (not directly observable) distribution of dark matter. Overdensities in the matter distribution are linked directly to initial density perturbations, the imprints of which we can also observe in the CMB. Statistically, these overdensities in the galaxy distribution (and equivalently of the matter field) can be described by the second-order correlation function and its Fourier space analogue, the power spectrum. These quantify the excess probability of finding two galaxies at a given separation with respect to a uniform distribution. To see why this statistic incorporates cosmological information, it is best to consider the most prominent feature: a series of peaks in the power spectrum called baryon-acoustic oscillations (BAOs)\,\cite{Eisenstein1998}. These peaks have the same physical origin as the oscillations of the CMB temperature power spectrum, namely the interplay between pressure and gravity for baryons before and during recombination. Therefore, the length scale associated with the BAOs is related to the sound horizon, stretched by the cosmic expansion. Thus, the BAO scale acts as a standard ruler and can be used to determine the cosmic expansion history.

To measure the correlation functions of galaxy clustering, we require three-dimensional positions. For these, spectroscopic redshifts are needed. Furthermore, to quantify dark energy's evolution, a wide redshift range extending beyond $z=1$ is needed. At these redshifts, though, many galaxies are mostly observable in the near-infrared (NIR). Thus, the extinction of Earth's atmosphere requires observations from space.

The second primary probe of the cosmic LSS is cosmic shear\,\cite{Kilbinger2015}\;\cite{Mandelbaum2018}. This effect describes the coherent distortion of the observed shapes of distant galaxies due to weak gravitational lensing by foreground matter structures. The lensing-induced shear in the galaxy images is a direct tracer of the projected matter distribution and, therefore, complementary to the galaxy clustering measurement, which depends on the bias between the galaxy and matter distribution.

For cosmic shear, again, two-point correlation functions (and the related power spectra) are the observable of choice. However, in contrast to galaxy clustering, one considers two-dimensional correlations of the shapes of galaxies at different projected positions on the sky. These correlations depend on the total projected mass between the observer and galaxies, as well as the distances between the observer, the lensing structure and the galaxies. The dependencies lead to the sensitivity of cosmic shear to the matter density, clustering, and cosmic expansion history. To optimally constrain the latter, the source galaxies need to be divided into tomographic redshift bins, and the correlations between different bins need to be considered. For this, redshift estimates for each galaxy need to be available. However, provided the bins are broader than associated errors, it is sufficient to use photometric instead of spectroscopic redshift estimates. This allows for a much larger source galaxy sample and extends the analysis to fainter, higher-redshift galaxies for which spectroscopic observations might be unfeasible.

The key sources of noise for cosmic shear analyses are the intrinsic shapes of galaxies. However, this noise is suppressed as more galaxies are considered, as the correlation function estimation averages over the randomly orientated shapes. Thus, for cosmic shear, it is vital to include as many galaxies as possible in one's sample.

\Euclid was specifically designed to meet the requirements of both galaxy clustering and cosmic shear. Thus, the start of this mission signals an exciting period for these two probes of the LSS.

\section{The \textit{Euclid} mission}

\Euclid is a medium-class mission of the European Space Agency (ESA) and part of its Cosmic Vision 2015-2025 programme. Historically, the mission arose from two proposed surveys, SPACE\,\cite{Cimatti2009}, whose goals were galaxy clustering measurements, and DUNE\,\cite{Refregier2009}, which planned a cosmic shear analysis. These surveys were proposed in 2007 and, due to their complementarity and similar goals, combined. Formal acceptance of the combined mission occurred in 2011. After a long design and development process to create the telescope and instruments to tight specifications, \Euclid finally launched on 1st July 2023. Launch and journey to the Earth-Sun Lagrange point 2 occurred as planned, and the first (unprocessed) test images were presented to the public less than a month later. The first processed early-release observations were released on 7th November 2023, highlighting the wide field-of-view (FOV), crisp resolution, and good detection of low-surface brightness objects. The scientific survey started on 14th February 2024, and the first cosmological results are expected in 2026.

As mentioned above, \Euclid's goal is to determine the evolution of dark energy. Of particular interest is the equation-of-state $w$ of dark energy, given by its pressure $p$ and density $\rho$ as $p=w\,\rho$. To cause the accelerated expansion of the Universe, the $w$ of dark energy needs to be smaller than $-1/3$. If $w$ is constant and equal to $-1$, dark energy behaves exactly as expected for a cosmological constant $\Lambda$. However, $w$ can be time-varying and thus a function of redshift $z$. Such a time-varying equation of state can be parameterized as
\begin{equation}
    w(z) = w_0 + w_a \frac{z}{1+z}\;,
\end{equation}
with free parameters $w_0$ and $w_a$. \Euclid aims to determine these parameters so precisely that the figure-of-merit (FOM), defined by
\begin{equation}
    \mathrm{FOM} = \frac{1}{\sqrt{\mathrm{det} \mathrm{Cov}(w_0, w_a)}}\;,
\end{equation}
reaches at least 400. Similarly, \Euclid is supposed to constrain $\sigma_8$ to 1\% and the sum of the neutrino masses $m_\nu$ to 0.03 eV. To reach these high goals, galaxy clustering and cosmic shear need to be combined.

For this, \Euclid uses a 1.2m primary mirror, which enables a large FOV of $0.54\, \mathrm{deg}^2$ and two scientific instruments, the optical visible imaging instrument VIS and the NIR spectrometer and photometer (NISP).

VIS\,\cite{Cropper2014}, specifically designed to observe large parts of the sky with a high resolution, consists of a large-format imager with a pixel size of $0\arcsecf1$. The large FOV is covered by 36 CCDs arranged in a square grid. VIS observes with a single bandpass $\IE$ in the range of 530--920 nm. This range was chosen because it contains the maxima of the spectral energy distribution of the targeted galaxies. Furthermore, galaxy shapes at these wavelengths are mostly regular, thus enabling the precise shape measurements required for cosmic shear analyses.

NISP\,\cite{Schirmer-EP18} combines a multiband NIR photometer and a slitless grism spectrometer. The photometer provides three passbands, $\YE$ (949.6\,nm--1212.3\,nm),  $\JE$(1167.6\,nm--1567\,nm) and $\HE$(1521.5\,nm--2021.4\,nm). Together, these can be used for photometric redshift estimates of galaxies whose shapes have been measured from VIS images. The spectrometer contains four grisms, which enable the simultaneous measurement of thousands of spectra across the FOV. Most observations are carried out with the red grisms (1206\,nm--1892\,nm), of which there are three in NISP with different dispersion directions. Two grims are used for science data acquisition in the EWS, allowing the spectroscopic redshifts $z$ of galaxies to be determined to an accuracy of $0.001(1+z)$. The NISP also carries a blue grism (926\,nm--1366\,nm), which is used to characterize the observed galaxy population and is important for non-cosmological science.

\begin{figure}
    \centering
    \includegraphics[width=0.7\linewidth]{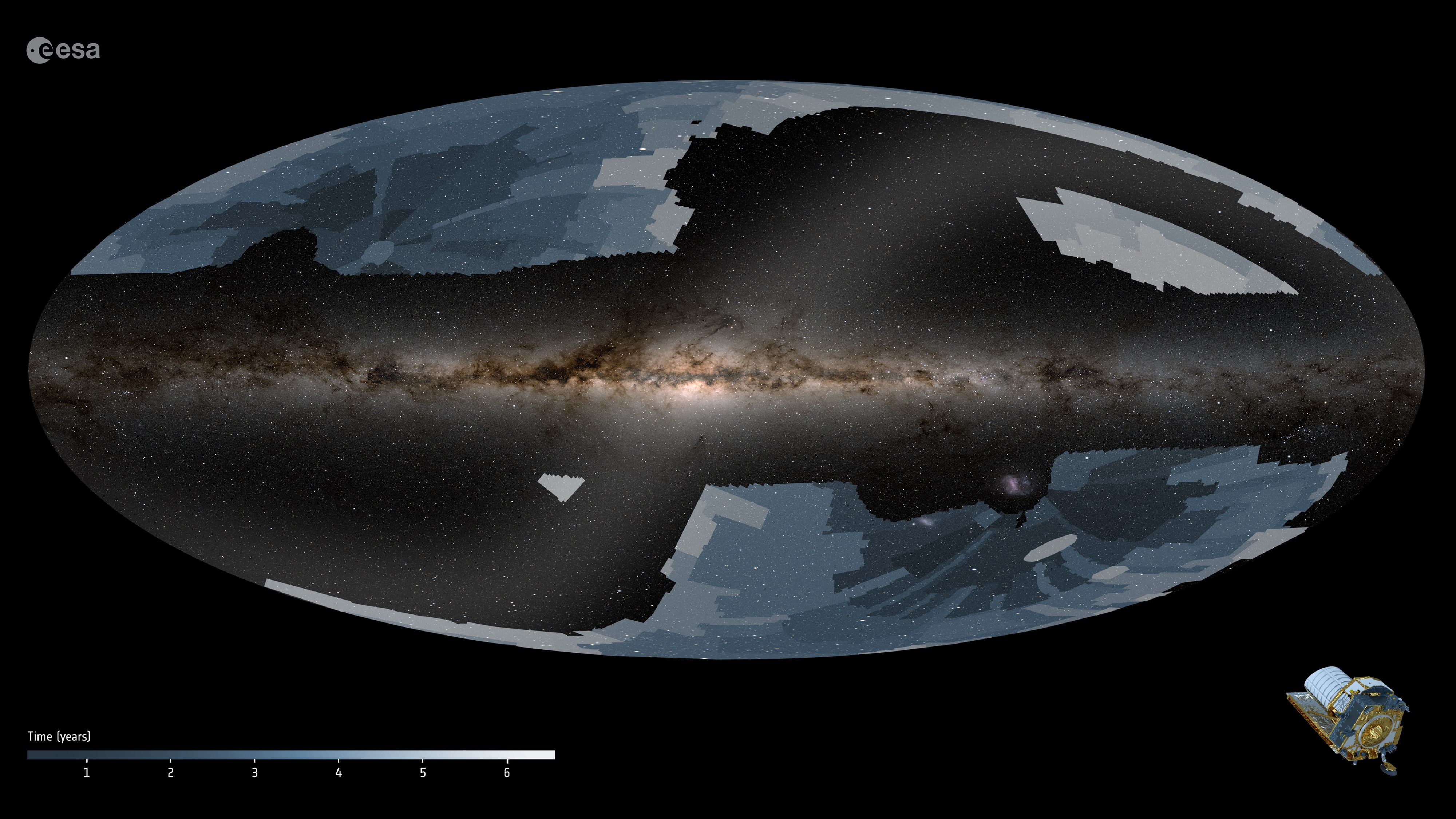}
    \caption{Area covered by the Euclid Wide Survey (EWS). Colours indicated in which survey year the area is observed. Image Credit: ESA/Euclid/Euclid Consortium. Work performed by ATG under contract for ESA}
    \label{fig: Euclid EWS}
\end{figure}

With these instruments, \Euclid will measure the shapes of 1.5 billion galaxies for cosmic shear analysis and the three-dimensional positions of 35 million galaxies for galaxy clustering analysis. To do so, \Euclid will undertake two surveys\,\cite{Scaramella-EP1}: the Euclid Wide Survey (EWS), shown in Fig.~\ref{fig: Euclid EWS}, and the Euclid Deep Survey (EDS).

The EWS will map up to $14\,000\, \mathrm{deg}^2$ of the extragalactic sky. This region encompasses almost all regions with low zodiacal background and low Galactic extinction. In this region, VIS will observe extended sources with an S/N $\geq 10$ up to a limiting magnitude of $\IE=24.5 \mathrm{mag}$; while NISP reaches an S/N$\geq 5$  up to a limiting magnitude of 24 mag in all three photometric bands. The EWS is divided into $27\,500$ fields, each of which is observed once by the roughly 70.2-minute reference observing sequence (ROS), which includes 6 VIS-exposures, simultaneously to  4 grism, 4 \YE-, 4 \JE- and 4 \HE-exposures, as well as bias, dark, and flat images. The EWS will provide the bulk of observing data for the cosmological analysis. However, to calibrate the telescope and instruments, characterise the observed galaxies, and enhance the galaxy shape measurements, it is accompanied by the EDS.

The EDS consists of three regions covering a total of $53\, \mathrm{deg}^2$. These fields are observed multiple times (at least 40 repetitions of the ROS), leading to limiting magnitudes fainter by two magnitudes than in the EWS. The goal of the EDS is to reach a 99\% complete and pure sample of $120\,000$ spectroscopically observed galaxies and accurate shapes of photometrically observed galaxies. These samples allow for the characterization of the EWS galaxy sample and calibration of the weak lensing shape measurement algorithms. The EDS regions were explicitly chosen to overlap with past and future observations, for example, the \textit{Herschel}\,\cite{Pearson2017} and AKARI-NEP surveys\,\cite{Lee2009}, the Chandra Deep Field South\,\cite{Treister2011},  \textit{Rubin} deep-drilling fields\,\cite{Ivezic2019}, and \textit{Spitzer}\,\cite{Moneti2022}, making them also a rich data set for non-cosmological science.

Additionally, \Euclid observes the Euclid Auxiliary Fields (EAFs), which are used to calibrate the photometric redshift estimation and quantify the colour dependence of shape measurements. The EAFs are chosen to have been observed already by ground- and space-based surveys and include, for example, the COSMOS field\;\cite{Scoville2007}.

\section{From images to cosmology}

The raw VIS and NISP images and spectroscopy measurements are processed in computing facilities on the ground to catalogues of galaxies with shape measures and redshift estimates, which are used to compute the correlation functions and finally infer the cosmological parameters.

As a first step, the VIS and NISP images are processed by combining dithers, considering flats and bias images, and extracting source catalogues. To complement these catalogues, external data from ground-based surveys like the Dark Energy Survey (DES)\,\cite{Abbott2016}, the Ultraviolet Near Infrared Optical Northern Survey (UNIONS), and the Legacy Survey of Space and Time (LSST)\,\cite{Ivezic2019} are calibrated and added to the \Euclid data set. These external data are crucial for estimating photometric redshift. Since VIS only observes in a single band, it cannot determine galaxies' redshifts. While NISP photometry provides colour information in the NIR, optical photometry is required to reach the tight constraints on the photometric redshift uncertainty.

Objects observed spectroscopically by NISP are assigned a spectroscopic redshift estimate. For these, the object position in the photometric image is mapped to its first-order dispersed image (the so-called spectrogram) in the spectroscopic image. The one-dimensional spectrum of the object is then extracted from the two-dimensional spectrogram. By template fitting to the spectrum, one then obtains the redshift estimate\,\cite{Paterson-EP32}.

For VIS-detected objects, shape measurements are carried out. These can be difficult, in particular for faint and small galaxies, as the intrinsic morphology of the galaxies is complex and unknown, instrument effects like the PSF, pixelisation, and detector noise distort the shapes, and blending or obscuration of galaxies further complicate the measurement. 

An important part of \Euclid's shape measurements is the correct understanding of the PSF. \Euclid forward-models the PSF by using our knowledge of the instrumental optics and including chromatic effects. However, this forward-model is not static, but updated and refined throughout the mission. These updates are based on empirical calibration against observed stars.


Once the shape and redshift estimates are added to the observed galaxy catalogue, summary statistics like the second-order correlation functions and power spectra are estimated. These are then compared to the cosmological model using Bayesian inference and MCMC techniques. While this might seem like a straightforward step, the high precision of \Euclid's measurements requires a high accuracy of theoretical models used for the \Euclid analysis. For example, modelling approximations accurate enough for previous analysis might no longer be appropriate for \Euclid. Some higher-order corrections to the matter power spectrum, such as magnification bias, can bias \Euclid cosmological parameter estimates by more than 1$\sigma$, if not taken into account \,\cite{Deshpande-EP28}. Consequently, we need to account for these effects in the modelling.
Additionally, astrophysical effects, even if accounted for in previous analyses, can play an even stronger role for \Euclid. An example of these are the correlations of intrinsic galaxy shapes, the so-called intrinsic alignments\,\cite{Joachimi2015}. These correlations contaminate the measured cosmic shear signal and, if not taken into account, can severely bias cosmological constraints\,\cite{Hirata2007}.
While previous cosmic shear analyses mostly used the simple phenomenological non-linear alignment model\,\cite{Bridle2007}, this might not be accurate enough at \Euclid precision and other models might be required\,\cite{Blazek2019}\;\cite{Fortuna2021}.

\begin{figure}
    \centering
    \includegraphics[width=\linewidth]{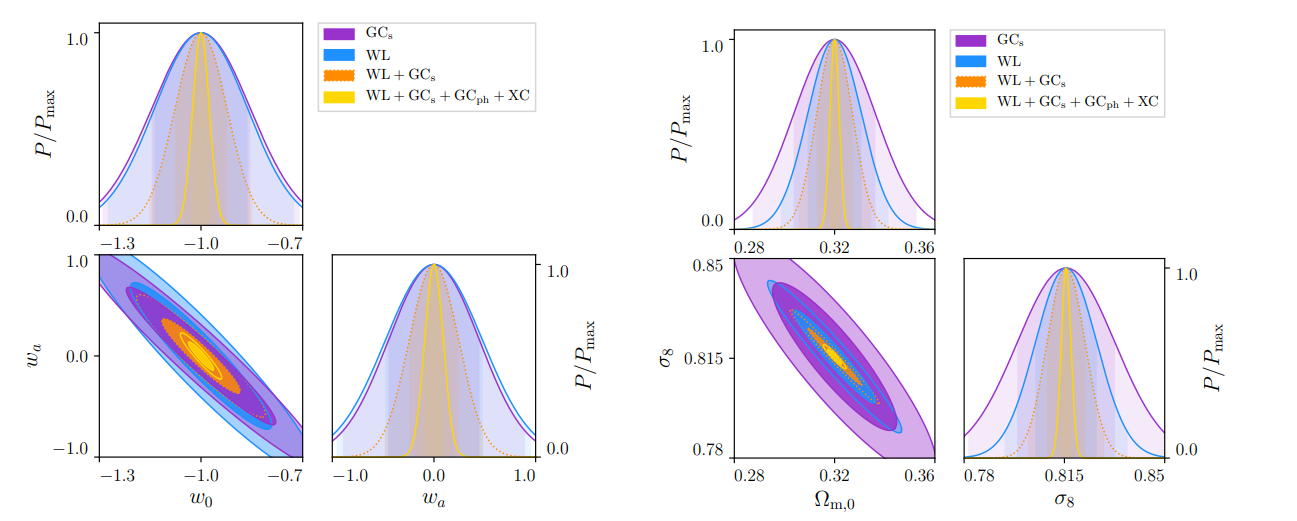}
    \caption[Forecast for \Euclid cosmological constraints]{Forecast for cosmological constraints from \Euclid. Purple are constraints using only galaxy clustering with spectroscopic redshift; yellow using only cosmic shear. Orange are constraints when combining spectroscopic clustering with cosmic shear and yellow from combining spectroscopic clustering, cosmic shear, galaxy clustering with photometric redshifts and the cross-correlations between clustering and shear. From Euclid Collaboration: Blanchard {\it et al.}\,\cite{Blanchard-EP7}}
    \label{fig:euclid forecast}
\end{figure}

Although these modelling concerns complicate the analysis, solving them will be worthwhile. Figure~\ref{fig:euclid forecast} shows a forecast of \Euclid constraints on a flat cosmology with varying $w_0$ and $w_a$, both from weak lensing and galaxy clustering alone and from their combination\,\cite{Blanchard-EP7}. One sees that the different probes are indeed complementary to each other. While lensing alone constrains $\Omega_\mathrm{m}$ up to 3.4\%, the full combination reaches 0.57\%. The full \Euclid dataset will constrain $w_0$ and $w_a$ to the per cent level. Thus \Euclid will definitively answer whether dark energy behaves like a cosmological constant or not.

\section{Additional cosmological probes}

While the design choices for the \Euclid mission were driven by galaxy clustering and cosmic shear power spectra, their cosmological information can be complemented by additional probes that benefit from the wide area, accurate shape measurements, and high resolution of \Euclid's surveys. 

Examples of these are higher-order statistics (HOS) of cosmic shear and galaxy clustering. These encompass statistical properties of the matter structures beyond the second-order correlation or the power spectra. HOS can be roughly grouped into two categories: those that consider $n$th-order moments of the density distribution (e.g. higher-order correlation functions\,\cite{Heydenreich2023}) and those using topological information (e.g. peak statistics\,\cite{Martinet2018}). These HOS depend differently an cosmological parameters than the standard power spectra. Thus, combining HOS and second-order statistics can partially resolve degeneracies and improve constraints on $\Omega_\mathrm{m}$ and $\sigma_8$ by factors between 2 and 5\,\cite{Ajani-EP29}.

Another important cosmological probe is the abundance of galaxy clusters. This abundance probes the clustering parameter $\sigma_8$, the matter density $\Omega_\mathrm{m}$, as well as the cosmic structure growth rate, which in turn is sensitive to $w_0$ and $w_a$. In the EWS, on the order of $10^6$ galaxy clusters of mass above $10^{14} M_\odot$ will be detected. Moreover, \Euclid will detect a large number of clusters at redshift beyond 1, which is critical to constraining dark energy. Aside from the direct detection of galaxy clusters, \Euclid is also particularly useful when combined with observation of the intra-cluster medium, e.g. from X-ray emission or the Sunyaev--Zeldovich (SZ) effect. These observations provide well-characterized cluster samples but lack estimates of the cluster masses. Here, weak lensing measurements from \Euclid will provide the necessary link between the observed X-ray luminosity or SZ-signal and the dark matter distribution.

Due to the fine spatial resolution and depth, up to $2300$ strongly lensed quasars are expected in the EDS. They provide an additional cosmological probe via time-delay measurements. Multiple images of the same source arrive at the observer at different times due to the lights' deflection and gravitational time delay. The differences between the images depend on cosmological distances and, therefore, the Hubble constant. Using the combination of strong and weak lensing, \Euclid's observations will also provide detailed mass models of galaxy clusters. These allow for constraining cosmological parameters and testing $\Lambda$CDM against alternative dark matter models.

Additional cosmological information can be accessed by correlating cosmic shear and the spatial galaxy distribution with maps of the gravitational lensing of the CMB. This CMB lensing is caused (partially) by the same LSS as the cosmic shear, but due to the large distance to the last scattering surface, the lensing signal depends differently on the cosmic expansion history and the structure growth rate. Similarly to HOS, CMB-LSS cross-correlations thus reduce parameter degeneracies. Combining \Euclid with data from the upcoming Simons Observatory can reduce the uncertainty on some parameters of generalised cosmological models by factors up to ten\,\cite{Ilic-EP15}.

\Euclid will observe $2\times 10^6$ quasars in the EWS, of which ten thousands fulfil the conditions for a tight relation between their rest-frame UV and rest-frame X-ray emission, i.e., unobscured UV and X-ray energies, no radio emission, and negligible host galaxy contribution\,\cite{Lusso2016}. These quasars can be calibrated to become standard candles and direct probes of the cosmic expansion history. By combining spectroscopic redshift estimates from NISP with X-ray data, $\Omega_\mathrm{m}$ and $w_0$ can be estimated independent of the primary cosmological probes.

Finally, \Euclid can also probe the cosmic expansion history with cosmic chronometers. These consist of massive, passively-evolving galaxies, which are a physically homogenous population and, at each redshift, are the oldest objects. Their formation history is well understood, so it is possible to determine their differential age as a function of redshift. This gives us an estimate of the relation between cosmic time and redshift and, thus, the cosmic expansion history. Forecasts for \Euclid expect to constrain $H_0$ with cosmic chronometers independently of other probes with a 4\% precision\,\cite{Moresco2022}.
 
\section{Non-cosmological science}

\Euclid data will also prove a treasure trove for a wide range of non-cosmological research.

For example, \Euclid will prove crucial for understanding the formation, evolution, and structure of galaxies. Due to its fine resolution, \Euclid will obtain resolved images of a considerable fraction of observed galaxies over a large redshift range. These unveil the morphologies of galaxies throughout cosmic time, giving clues to galaxy evolution models. \Euclid will help in characterizing physical properties like the star-formation rate (SFR) of galaxies throughout time by observing up to 67 million H$\alpha$-emitting galaxies up to $z=1.8$ in the EWS\,\cite{Pozzetti2016}. As the H$\alpha$ - emission intensity is strongly correlated to the SFR, these observations constrain the SFR-to-stellar mass relation. Moreover, due to its ability to detect low-surface brightness objects, \Euclid will observe many faint features of nearby galaxies, such as stellar streams or tidal remnants, as well as dwarf galaxies\,\cite{Borlaff-EP16}. These objects are crucial for understanding galaxy formation at the low-mass end. Finally, \Euclid can also reveal the relation between galaxy properties and the dark matter halos they reside in. For this, the weak lensing measurements can be used to trace the dark matter distribution, which is then compared to the position of galaxies with varying physical properties. These correlations are crucial tests of galaxy evolution models.


\section{Conclusion}
\Euclid is a crucial experiment for exploring the dark universe, particularly the time evolution of dark energy, by tracing the cosmic large-scale structure. It is designed, built, and calibrated to deliver some of the most precise galaxy clustering and cosmic shear results ever obtained.

After its successful launch on 1st July 2023, travel to L2, and science performance verification, the scientific survey started on 14th February 2024, with an expected mission length of 6 years. At the time of writing, the first data are pouring in, and the first science results are prepared to be published. The first cosmological results will be published in 2026, together with a full release of all data observed in the first survey year.

Several additional probes of cosmology complement the primary cosmological probes. This is useful for achieving the best cosmological results as well as for checking and verifying consistency. \Euclid's surveys will also provide an excellent dataset for non-cosmological research, including galaxy evolution and stellar physics.

\section*{Acknowledgments}
LL acknowledges support from the Austrian Research Promotion Agency (FFG) and the Federal Ministry of the Republic of Austria for Climate Action, Environment, Mobility, Innovation and Technology (BMK) via grant 900565.
\AckEC

\section*{References}

\bibliography{biblio}

\begin{thebibliography}{10}

\bibitem{Planck2020_VI}
{Planck Collaboration: N. Aghanim {\it et al.}}
\newblock {\em \aap}, 641:A6, 2020.

\bibitem{DES2023}
{Dark Energy Science Collaboration: T.~Abbott {\it et al.}}
\newblock {\em \PRD}, 107(8):083504, 2023.

\bibitem{Troester2021}
T.~{Tr{\"o}ster {\it et al.}}
\newblock {\em \aap}, 649:A88, 2021.

\bibitem{Riess2022}
A.~G. {Riess {\it et al.}}
\newblock {\em \apjl}, 934(1):L7, 2022.

\bibitem{Abdalla2022}
E.~{Abdalla {\it et al.}}
\newblock {\em Journal of High Energy Astrophysics}, 34:49--211, 2022.

\bibitem{Laureijs2011}
R.~{Laureijs {\it et al.}}
\newblock arXiv:1110.3193, 2011.

\bibitem{Eisenstein1998}
D.~J. {Eisenstein} and W.~{Hu}.
\newblock {\em \apj}, 496(2):605--614, 1998.

\bibitem{Kilbinger2015}
M.~{Kilbinger}.
\newblock {\em Reports on Progress in Physics}, 78(8):086901, 2015.

\bibitem{Mandelbaum2018}
R.~{Mandelbaum}.
\newblock {\em \araa}, 56:393--433, 2018.

\bibitem{Cimatti2009}
A.~{Cimatti {\it et al.}}
\newblock {\em Experimental Astronomy}, 23(1):39--66, 2009.

\bibitem{Refregier2009}
A.~{Refregier}.
\newblock {\em Experimental Astronomy}, 23(1):17--37, 2009.

\bibitem{Cropper2014}
M.~{Cropper {\it et al.}}
\newblock In {\em Space Telescopes and Instrumentation 2014: Optical, Infrared, and Millimeter Wave}, volume 9143 of {\em Society of Photo-Optical Instrumentation Engineers (SPIE) Conference Series}, page 91430J, 2014.

\bibitem{Schirmer-EP18}
{Euclid Collaboration: M.~Schirmer {\it et al.}}
\newblock {\em \aap}, 662:A92, 2022.

\bibitem{Scaramella-EP1}
{Euclid Collaboration: R.~Scaramella {\it et al.}}
\newblock {\em \aap}, 662:A112, 2022.

\bibitem{Pearson2017}
C.~{Pearson \textit{et al.}}
\newblock {\em Publication of Korean Astronomical Society}, 32(1):219--223, 2017.

\bibitem{Lee2009}
H.~M. {Lee \textit{et al.}}
\newblock {\em \pasj}, 61:375, 2009.

\bibitem{Treister2011}
E.~{Treister \textit{et al.}}
\newblock {\em \nat}, 474(7351):356--358, 2011.

\bibitem{Ivezic2019}
{\v{Z}}.~{Ivezi{\'c} {\it et al.}}
\newblock {\em \apj}, 873(2):111, 2019.

\bibitem{Moneti2022}
{Euclid Collaboration: A.~Moneti \textit{et al.}}
\newblock {\em \aap}, 658:A126, 2022.

\bibitem{Scoville2007}
N.~{Scoville \textit{et al.}}
\newblock {\em \apjs}, 172(1):1--8, 2007.

\bibitem{Abbott2016}
{Dark Energy Survey Collaboration: T.~Abbott {\it et al.}}
\newblock {\em \PRD}, 94(2):022001, 2016.

\bibitem{Paterson-EP32}
{Euclid Collaboration: K.~Paterson {\it et al.}}
\newblock arXiv:2303.15525, 2023.

\bibitem{Deshpande-EP28}
{Euclid Collaboration: A.~C.~Deshpande {\it et al.}}
\newblock arXiv:2302.04507, 2023.

\bibitem{Joachimi2015}
B.~{Joachimi {\it et al.}}
\newblock {\em \ssr}, 193(1-4):1--65, 2015.

\bibitem{Hirata2007}
C.~M. {Hirata {\it et al.}}
\newblock {\em \mnras}, 381(3):1197--1218, 2007.

\bibitem{Bridle2007}
S.~{Bridle} and L.~{King}.
\newblock {\em New Journal of Physics}, 9(12):444, 2007.

\bibitem{Blazek2019}
J.~A. {Blazek {\it et al.}}
\newblock {\em \PRD}, 100(10):103506, 2019.

\bibitem{Fortuna2021}
M.~C. {Fortuna {\it et al.}}
\newblock {\em \aap}, 654:A76, 2021.

\bibitem{Blanchard-EP7}
{Euclid Collaboration: A.~Blanchard {\it et al.}}
\newblock {\em \aap}, 642:A191, 2020.

\bibitem{Heydenreich2023}
S.~{Heydenreich {\it et al.}}
\newblock {\em \aap}, 672:A44, 2023.

\bibitem{Martinet2018}
N.~{Martinet {\it et al.}}
\newblock {\em \mnras}, 474(1):712--730, 2018.

\bibitem{Ajani-EP29}
{Euclid Collaboration: V.~Ajani {\it et al.}}
\newblock {\em \aap}, 675:A120, July 2023.

\bibitem{Ilic-EP15}
{Euclid Collaboration: S. Ili{\'c} {\it et al.}}
\newblock {\em \aap}, 657:A91, 2022.

\bibitem{Lusso2016}
E.~{Lusso} and G.~{Risaliti}.
\newblock {\em \apj}, 819(2):154, 2016.

\bibitem{Moresco2022}
M.~{Moresco {\it et al.}}
\newblock {\em Living Reviews in Relativity}, 25(1):6, 2022.

\bibitem{Pozzetti2016}
L.~{Pozzetti {\it et al.}}
\newblock {\em \aap}, 590:A3, 2016.

\bibitem{Borlaff-EP16}
{Euclid Collaboration: A.~S.~Borlaff {\it et al.}}
\newblock {\em \aap}, 657:A92, 2022.

\end{thebibliography}







\end{document}